\documentclass[a4paper,10pt]{article}
\usepackage[utf8x]{inputenc}

\title{ Harmonic Superspace Gaugeon Formalism for the ABJM Theory }
\author{ Mir Faizal \\
Mathematical Institute, University of Oxford
\\ Oxford
OX1 3LB, United Kingdom 
 }

\begin{document}

\maketitle

\begin{abstract}
In this paper we will 
 analyse the 
 ABJM  theory   in  harmonic superspace.
The harmonic superspace variables will be
 parameterized by the coset $SU(2)/U(1)$ and thus will
have manifest $\mathcal{N} =3$ supersymmetry. 
 We  will study the 
 quantum gauge transformations  and the BRST transformations
of this theory in gaugeon formalism. We will use this BRST symmetry to project out the physical 
sub-space from the total Hilbert space. We will also show 
that the evolution of the $\mathcal{S}$-matrix is unitary for this 
ABJM  theory   in   harmonic superspace.
\end{abstract}

\section{Introduction}
The harmonic superspace
 is parametrized by the coset $SU(2)/U(1)$ and thus has  
 $\mathcal{N} =2$ supersymmetry in four dimensions
  \cite{h1}-\cite{h2}, 
and $\mathcal{N}= 3$ supersymmetry in three dimensions 
\cite{h21}-\cite{h4}.   A superconformal Chern-Simons-matter theory called the ABJM theory   has already 
been studied  in this Harmonic superspace \cite{ahs}.
The ABJM theory that 
is thought to capture the dynamics of  multiple $M2$-branes. 
It  has manifest
$\mathcal{N} = 6$ supersymmetry which is expected to get
enhanced to $\mathcal{N} = 8$ supersymmetry \cite{su}-\cite{0su1}.
The ABJM theory is invariant under 
  the gauge group 
$U(N)_k \times U(N)_{-k}$
   \cite{apjm}. 
As the ABJM theory has a gauge symmetry associated with it, we have to fix a gauge in order to
quantization  it. The gauge fixing condition can be incorporated at a quantum level 
by adding  ghost and gauge fixing terms to the original classical Lagrangian density. 
 However,  
the Fock space in one gauge  is  different
 from that in other gauge because 
the Fock space in any  gauge  
is not wide enough to realize the quantum gauge freedom.
This problem is solved by 
 the gaugeon formalism \cite{1l}-\cite{2l}. 
In this formalism 
 the quantum gauge transformation are accounted for by introducing 
 a set of extra fields called 
gaugeon fields.  
As the ABJM 
theory has gauge symmetry associated with it, so it
can be analysed in the gaugeon formalism. This has already been 
done for ABJM theory in $\mathcal{N} =1$ superspace formalism
\cite{ne1}. In this paper we will do it for ABJM theory 
in $\mathcal{N} =3 $ harmonic superspace. It may be noted that unlike
the ABJM theory in $\mathcal{N} =1$ superspace, the ABJM theory
in $\mathcal{N} =3 $ harmonic superspace contains no 
explicit superspace potential term. 
We will also study the extended BRST symmetry of this model and use it
to show that the $\mathcal{S}$-matrix for the ABJM
theory in $\mathcal{N} =3 $ harmonic superspace is unitarity. 
It may be noted that similar  results could be obtained by using  conventional BRST transformations
along with the  Yokoyama's subsidiary 
condition \cite{y2a}-\cite{ya1}.

The BRST symmetry for ordinary Chern-Simons theory has  been  throughly studied
\cite{16}-\cite{17}. In fact,   the BRST symmetry of $\mathcal{N} = 1$  
  Chern-Simons theory   has also been analysed \cite{18, 19}. 
The BRST of the Chern-Simons theory is similar to 
the BRST symmetry of Yang-Mills theories. This is because the structure of 
the gauge fixing term and the ghost terms is similar in both these theories. The BRST 
symmetry  for gauge theories has also been studied in the background field method.  In the 
background field method,  all the fields in a theory  are shifted.  
The BRST 
symmetry of these shifted fields can be 
analysed by the use of 
Batalin-Vilkovisky   formalism \cite{10a}-\cite{11a}.  
In this formalism the a modified  BRST symmetry arise  due to the
 invariance of a
 theory under the original BRST 
transformations along with 
these shift transformation. This  
    have been done for the ordinary 
Yang-Mills theories and the ordinary  
Chern-Simons theories \cite{12a}-\cite{13a}. 
Furthermore, as the BRST transformations 
 mix
 the fermionic and bosonic 
coordinates, they can be viewed as  supersymmetric transformations. 
The BRST transformations of gauge theories  have been expressed 
  in the extended 
superspace formalism by adding new Grassmann 
coordinates~\cite{14a}-\cite{30}. The main use of the BRST symmetry is to project 
out the physical sub-space from 
the total Hilbert space of the theory. Thus, in this paper we will analyse the BRST symmetry for 
the ABJM theory and use it to project out the physical sub-space from the total Hilbert space of the theory.

\section{Harmonic superspace}
In this paper we use the
harmonic variables $u^{\pm}$  subjected to the 
constraints 
\begin{eqnarray}
 u^{+i} u^-_i = 1, && u^{+i} u^+_i = u^{-i} u^-_i =0. 
\end{eqnarray}
These harmonic variable parameterize the coset $SU(2)/U(1)$. 
So, this superspace is parameterized by the following coordinates 
\begin{equation}
 z = ( x^{ab}, \theta_{a}^{++}, \theta^{--}_a, \theta^0_a, u_i^{\pm} ),
\end{equation}
where $ \theta^{\pm}_a = \theta_a^{ij} u^{\pm}_i u^{\pm}_j$ 
and $\theta^0_a = \theta^{ij}_a u^+_i u^-_j$. 
Now we can define the following derivatives
\begin{eqnarray}
\nonumber \\ 
{ 
D}^{++}&=&\partial^{++}+2i\theta^{++ a}\theta^{0 b}
 \partial^A_{ab}
 +\theta^{++a}\frac\partial{\partial\theta^{0 a}}
 +2\theta^{0 a}\frac\partial{\partial\theta^{--a}},\nonumber \\
{  D}^{--}&=&\partial^{--}
 -2i\theta^{--a}\theta^{0 b}\partial^A_{ab}
 +\theta^{--a}\frac\partial{\partial\theta^{0 a}}
 +2\theta^{0 a}\frac\partial{\partial\theta^{++ a}},
\nonumber \\
{  D}^0&=&\partial^0+2\theta^{++ a}\frac\partial{\partial\theta^{++ a}}
-2\theta^{--a}\frac\partial{\partial\theta^{--a}}, 
\end{eqnarray}
and
\begin{eqnarray}
D^{--}_a=\frac\partial{\partial\theta^{++ a}}
 +2i\theta^{--b}\partial^A_{ab}, &&
D^0_a= -\frac12\frac\partial{\partial\theta^{0 a}}
+i\theta^{0 b}\partial^A_{ab},\nonumber \\ 
D^{++}_a=\frac{\partial}{\partial
\theta^{--a}}, &&
\end{eqnarray}
where 
\begin{eqnarray}
&\partial^{++}=u^+_i\frac\partial{\partial u^-_i}, &
\partial^{--}=u^-_i\frac\partial{\partial u^+_i},\nonumber \\ 
&\partial^0 = u^+_i\frac\partial{\partial u^+_i}
-u^-_i\frac\partial{\partial u^-_i}.& 
\end{eqnarray}
These derivatives are satisfy 
\begin{eqnarray}
 \{D^{++}_a, D^{--}_b\}=2i\partial^A_{ab}, \quad \{D^{0}_a,
D^{0}_b\}=-i\partial^A_{ab}, 
\nonumber \\ 
{[{ D}^{\mp\mp}, D^{\pm\pm}_a]}=2D^0_a, \quad [{  D}^{0},
D^{\pm\pm}_a]=\pm 2D^{\pm\pm}_a, 
\nonumber \\ 
\partial^0=[\partial^{++},\partial^{--}],\quad
[{  D}^{++}, {  D}^{--}]={  D}^0. \nonumber \\
\{D^{\pm\pm}_a, D^{0}_b\} = 0\,,\quad [{ 
D}^{\pm\pm}, D^0_a]=D^{\pm\pm}_a.
  \end{eqnarray}

The analytic
superfields $\Phi_A = \Phi_A(\zeta_A)$ are  independent of the
$\theta^{--}_a$, thus defined by $ D^{++}_a\Phi_A=0$.
The analytic subspace 
is parametrized by the
following coordinates
\begin{eqnarray}
\zeta_A=(x^{ab}_A,
\theta^{++}_a, \theta^{0}_a, u^\pm_i), 
  \end{eqnarray}
where
\begin{eqnarray}
x^{ab}_A=(\gamma_m)^{ab}x^m_A=x^{ab}
+i(\theta^{++a}\theta^{--b}+\theta^{++b}\theta^{--a}).
  \end{eqnarray}
The generators of the supersymmetry are given by 
\begin{eqnarray}
&Q^{++}_a=u^+_iu^+_j Q_a ^{ij}, &
Q^{--}_a=u^-_iu^-_j Q_a ^{ij},\nonumber \\ 
&Q^0_a = u^+_iu^-_j Q_a ^{ij},& 
\end{eqnarray}
where 
\begin{equation}
 Q_a^{ij} = \frac{\partial}{\partial \theta^a_{ij}} - \theta^{ijb } 
 \partial_{ab}.
\end{equation}
It is convenient to denote the superspace measure in this superspace as 
  \begin{eqnarray}
d^9z &=&-\frac1{16}d^3x
(D^{++})^2 (D^{--})^2(D^{0})^2, \nonumber \\
d\zeta^{(-4)}&=&\frac{1}{4} d^3x_Adu (D^{--})^2(D^{0})^2\,.
  \end{eqnarray}
A conjugation in the $\mathcal{N} =3$ harmonic superspace is defined  by
  \begin{eqnarray}
\widetilde{(u^\pm_i)}=u^{\pm i},\quad \widetilde{(x^m_A)}=x^m_A,
 \nonumber \\  \widetilde{(\theta^{\pm\pm}_a)}=
\theta^{\pm\pm}_a,\quad \widetilde{(\theta^0_a)}=
\theta^0_a.
  \end{eqnarray}
It is squared to $-1$ on the harmonics  and to $1$ on $x^m_A$  and 
 Grassmann coordinates. So,  the analytic superspace measure is real
 $\widetilde{d\zeta^{(-4)}}=d\zeta^{(-4)}$
and the full superspace measure is imaginary $\widetilde{d^9z}=-d^9z$. 
\section{ABJM Theory}
Now we can construct the Lagrangian density  for  ABJM theory in this deformed 
superspace using $V^{++}_L$ and $V^{++}_R$, which are defined by
\begin{eqnarray}
V^{++}_L &=& u^+_i u^+_j V^{ij}_L, \nonumber \\ 
V^{++}_R &=& u^+_i u^+_j V^{ij}_R,
\end{eqnarray}
where $V^{ij}_L$ and
$V^{ij}_R$ are  fields transforming under the gauge group 
 $U(N)_k$ and $U(N)_{-k}$,  respectively. 
We also define matter fields $q^{+}$ and $\bar q^{+}$, 
which transform under the bifundamental representation 
of the group   $U(N)_k \times U(N)_{-k}$.
The Lagrangian density  for the  ABJM theory can now be written as 
\begin{equation}
\mathcal{L}_{ABJM} = \mathcal{L}_{CS, k} [ V^{++}_L]
  + \mathcal{L}_{CS, - k} [ V^{++}_R]
   + \mathcal{L}_{M} [ q^{+}, \bar q^{+}] ,
\end{equation}
where 
\begin{eqnarray}
 \mathcal{L}_{CS, k}[ V^{++}_L]  &=&\frac{ik}{4\pi}\, 
tr\sum\limits^{\infty}_{n=2}
 \frac{(-1)^{n}}{n} \int
 d^6\theta du_{1}\ldots du_n  H^{++}_L, \nonumber \\
  \mathcal{L}_{CS, -k}[ V^{++}_R] &=&- 
\frac{ik}{4\pi}\,tr\sum\limits^{\infty}_{n=2} 
\frac{(-1)^{n}}{n} \int
d^6\theta du_{1}\ldots du_n H^{++}_R, \nonumber \\
\mathcal{L}_{M} [ q^{+}, \bar q^{+}] 
 &=&tr\int d\zeta^{(-4)}\bar q^{+}  \nabla^{++}
   q^{+},
\end{eqnarray}
and 
\begin{eqnarray}
 H^{++}_L &=& \frac{V^{++}(z,u_{1} )_L   V^{++}(z,u_{2} )_L\ldots
  V^{++}(z,u_n )_L }{ (u^+_{1} u^+_{2})\ldots (u^+_n u^+_{1} )},
\nonumber \\ 
H^{++}_R &=& \frac{V^{++}(z,u_{1} )_R   V^{++}(z,u_{2} )_R\ldots
  V^{++}(z,u_n )_R }{ (u^+_{1} u^+_{2})\ldots (u^+_n u^+_{1} )}, 
\nonumber \\ 
\nabla^{++}q^{+}&=&{  D}^{++}q^{+}
 + V^{++}_L  q^{+}- q^{+}  V^{++}_R, 
 \nonumber \\   \nabla^{++}\bar q^{+}&=&{  D}^{++}\bar q^{+}
 -\bar q^{+}   V^{++}_L +  V^{++}_R   \bar q^{+}.
\end{eqnarray}
The covariant derivatives for the matter fields are given by 
\begin{eqnarray}
\nabla^{++}q^{+}&=&{  D}^{++}q^{+}
 + V^{++}_L  q^{+}- q^{+}  V^{++}_R\,,  \nonumber \\  
 \nabla^{++}\bar q^{+}&=&{  D}^{++}\bar q^{+}
 -\bar q^{+}   V^{++}_L +  V^{++}_R   \bar q^{+},
\end{eqnarray}
It is useful to define $V^{--}_L$ and $ V^{--}_R$ as
\begin{eqnarray}
 V^{--}_L&=&\sum_{n=1}^\infty (-1)^n \int du_1\ldots
du_n  E^{++}_L, \nonumber \\ 
 V^{--}_R&=& \sum_{n=1}^\infty (-1)^n \int du_1\ldots
du_n  E^{++}_R,
\end{eqnarray}
where 
\begin{eqnarray}
  E^{++}_L &=& \frac{V^{++}(z,u_1) V^{++}_L(z,u_2)\ldots  
V^{++}_L(z,u_n)}{(u^+u^+_1)(u^+_1u^+_2)\ldots (u^+_n u^+)}, \nonumber \\
 E^{++}_R &=& \frac{V^{++}(z,u_1) V^{++}_R(z,u_2)\ldots  
V^{++}_R(z,u_n)}{(u^+u^+_1)(u^+_1u^+_2)\ldots (u^+_n u^+)}.
\end{eqnarray}
It is also useful to define $ W^{++}_L$ and $ W^{++}_R $ as 
\begin{eqnarray}
  W^{++}_L &=& -\frac{1}{4} D^{++a} D^{++}_{ a}  V^{--}_L, \nonumber \\ 
 W^{++}_R &=& -\frac{1}{4} D^{++a} D^{++}_{ a}  V^{--}_R. 
\end{eqnarray}
This ABJM theory is invariant under the following $\mathcal{N}= 3$ supersymmetric transformations,  
\begin{eqnarray}
\delta_\epsilon q^{+}&=& i\epsilon^{a}\hat\nabla^0_a  q^{+}\,,
\nonumber \\
\delta_\epsilon\bar q^{+} &=&i\epsilon^{a} \hat\nabla^0_a  
\bar q^{+ }\,, \nonumber  \\
\delta_\epsilon V^{++}_L&=&\frac{8\pi}k\epsilon^{a}
 \theta^0_a  q^+\bar q^+\,, \nonumber \\
\delta_\epsilon V^{++}_R &=&\frac{8\pi}k\epsilon^{a}
 \theta^0_a  \bar q^+  q^+\,,
\end{eqnarray}
where
\begin{eqnarray}
 \hat\nabla^0_a  q^{+}& =& \nabla^0_a  q^{+} 
+\theta^{--}_a (W^{++}_L  q^{+} -q^{+}  W^{++}_R )\,, \nonumber \\
 \nabla^0_a  q^+&=&D^0_a q^+
 +V^0_{L\, a} q^+ -q^+ V^0_{R\,a }\,, 
\nonumber \\ V^0_{L\, a}&=&-\frac12D^{++}_a
V^{--}_{L},   \nonumber \\ 
 V^0_{R\, a}&=&-\frac12D^{++}_a
V^{--}_{R}.
\end{eqnarray}
Thus, apart from the original manifest $\mathcal{N} =3$ 
supersymmetry, this model has additional $\mathcal{N} =3$ supersymmetry, $\delta_\epsilon \mathcal{L}_{ABJM} =0 $. 
So, this ABJM theory has $\mathcal{N} =6$ supersymmetry. 
\section{Gaugeon Formalism}
 In the gaugeon formalism  extra fields are added to account for  the 
 quantum gauge transformations. Thus, apart from the gauge fixing and 
ghost terms a gaugeon term is also added to the original classical Lagrangian density.
In this section we will analyse the ABJM theory in gaugeon formalism. 
This ABJM theory is invariant under the  infinitesimal gauge
 transformations, $\delta \mathcal{L}_{ABJM} =0$,  
\begin{eqnarray}
\delta q^{+} &=& \Lambda_L   q^{+}-q^{+}  \Lambda_R,\nonumber \\
 \delta\bar q^{+} &=&\Lambda_R   \bar q^{+}-\bar q^{+}  \Lambda_L,
\nonumber \\
\delta V^{++}_L&=&\nabla^{++} \Lambda_L,\nonumber \\
\delta V^{++}_R&=&\nabla^{++} \Lambda_R,  
\end{eqnarray}
where 
\begin{eqnarray}
\nabla^{++}  \Lambda_L&=&-{  D}^{++}\Lambda_L -[V^{++}_L,\Lambda_L] ,
\nonumber \\
\nabla^{++}  \Lambda_R&=&-{  D}^{++}\Lambda_R -[V^{++}_R,\Lambda_R].  
\end{eqnarray}
Due to the invariance of the ABJM theory under these infinitesimal gauge transformations,
 we can not quantize it without fixing a gauge. 
So, we choose the following gauge fixing conditions,  
\begin{eqnarray}
D^{++}     V_L^{++} =0, && D^{++}      V^{++}_R =0. 
\end{eqnarray}
To incorporate these   gauge fixing conditions  at 
the quantum level we have to add the corresponding gauge fixing and 
ghost terms to the original classical Lagrangian density. In order to incorporate the  
 quantum gauge transformations we have to also add 
a gaugeon term to it. 
Thus, the  total Lagrangian  density
 is given  by the sum of the original classical Lagrangian density $ \mathcal{L}_{ABJM}$,
 the gauge fixing 
term $ \mathcal{L}_{gf}$, the ghost term $ \mathcal{L}_{gh}$, and the gaugeon term $ \mathcal{L}_{go}$, 
\begin{equation}
 \mathcal{L}_t =  
 \mathcal{L}_{ABJM} +   \mathcal{L}_{gf} +   \mathcal{L}_{gh} +
  \mathcal{L}_{go},
\end{equation}
where 
\begin{eqnarray}
\mathcal{L}_{gf} &=& \int  d\zeta^{(-4)} tr  \left[b_L    
 (D^{++}V^{++}_L) + \frac{\alpha}{2}b^2_L - \frac{\alpha}{2}\tilde{b}^2_R     \right. \nonumber \\ && \left.
\,\,\,\,\,\,\,\,\,\,\,\,\,\,\,\,\,\,\,\,\,\,\,\,\,\,\, \,\, -
{b}_R      (D^{++}V^{++}_R)   
\right]_|, 
\nonumber \\ 
\mathcal{L}_{gh} &=& \int d\zeta^{(-4)} tr 
\left[ \overline{c}_L      D^{++}\nabla^{++}      c_L - \overline{c}_R 
     D^{++}\nabla^{++}   
  c_R \right]_|.
\nonumber \\ 
 \mathcal{L}_{go} &=& \int d\zeta^{(-4)} 
tr \left[D^{++}\overline{y}_L D^{++}y_L + \frac{1}{2}
 (\overline{y}_L  + \alpha b_L)^2 \right. \nonumber \\ && \left.
\,\,\,\,\,\,\,\,\,\,\,\,\,\,\,\,\,\,\,\,\,\,\,\,\,\,\, \,\,- 
D^{++}\overline{k}_L D_a k 
 - D^{++}\overline{y}_R D^{++} y_R \right. \nonumber \\ && \left.
\,\,\,\,\,\,\,\,\,\,\,\,\,\,\,\,\,\,\,\,\,\,\,\,\,\,\,\,\,- \frac{1}{2}
 (\overline{y}_R + \alpha  b_R)^2  + D^{++}\overline{k}_R 
D_a  k_R
   \right]_|.
\end{eqnarray}

Now we can analyse the quantum gauge transformations
by varying the gauge fixing parameter $\alpha$.  
So, we transform $\alpha $ as 
\begin{equation}
 \delta_{qg}\, \alpha = \tau \alpha.
\end{equation}
Under this transformation the  matter fields transform  as, 
\begin{eqnarray}
 \delta_{qg}\, q = i(\tau \alpha  y  q_L - q \tau \alpha   y_R ), 
 &&   \delta_{qg}\, \bar q = i(  \tau \alpha   y_R 
\bar q  - \bar q \tau \alpha  y_L).
\end{eqnarray}
and all other fields transform as, 
\begin{eqnarray}
 \delta_{qg}\, V^{++}_L =  \tau\nabla^{++}( \alpha y_L ),  && 
 \delta_{qg}\, V^{++}_R =  \tau \nabla^{++}( \alpha y_R ), \nonumber \\
 \delta_{qg}\, \overline y_L = \tau \alpha b_L  && 
 \delta_{qg}{\overline{y}}_R =\tau \alpha
  b_R,  \nonumber \\
 \delta_{qg}\, c_L =  [\tau c_L,  \alpha y_L]  + \tau \alpha k_L, && 
 \delta_{qg}\, c_R =  [\tau  c_R, \alpha  y_R]  + \tau \alpha  k_R,
 \nonumber \\
 \delta_{qg}\, \overline{c}_L =   [\tau \overline{c}_L, \alpha y_L], &&
 \delta_{qg}\, {\overline{c}}_R =   [\tau {\overline{c}}_R, \alpha {y}_R],
 \nonumber \\
 \delta_{qg}\, \overline{k}_L = - \tau \alpha c_L, &&
  \delta_{qg}\, {\overline{k}}_R = - \tau \alpha {c}_R, \nonumber \\
 \delta_{qg}\, b_L =  [\tau b_L, \alpha y_L] - 
[\tau \overline{c}_L, \alpha k_L], &&
 \delta_{qg}\,  b_R =  [\tau  b_R, \alpha y_R] - 
[\tau {\overline{c}}, \alpha  k], 
\nonumber \\  
 \delta_{qg}\, y_L = \delta_{qg}\, k_L  = 0, 
&&  \delta_{qg}\,  y_R=  \delta_{qg}\,  k_R  = 0.  
\end{eqnarray}
The total Lagrangian  density
is invariant under these  quantum    transformations, 
$ \delta_{qg}\, \mathcal{L}_t =0$. 
Thus, by adding a gaugeon term
 we have derived the quantum gauge transformation 
for the ABJM theory in harmonic superspace. 

\section{BRST Symmetry}
The ABJM theory was initially invariant under gauge transformations. But, 
after fixing a gauge this gauge invariance was broken. However, the sum 
of the original Lagrangian density,  gauge fixing term, the ghost term and  
 the gaugeon term is invariant under a new symmetry called the BRST symmetry. The BRST 
transformations of the matter fields are given by 
\begin{eqnarray}
 s \, q = i(c_L q - q c_R),  
&&  s \, \bar q = i(  c_R \bar q - \bar q c_L).
\nonumber \\
\end{eqnarray}
and the BRST transformations of all other fields are given by 
\begin{eqnarray}
s \,V^{++}_L= \nabla^{++}      c_L, && s\, V^{++}_R =\nabla^{++} 
     c_R, \nonumber \\
s \,c_L = - \frac{1}{2}{\{c_L, c_L\}}_ {     } , && s \,\overline{c}_R = 
 b_R, \nonumber \\
s \,\overline{c}_L = b_R, && s \, c_R = 
- \frac{1}{2}\{\tilde c_R, \tilde c_R\}_{     }, \nonumber \\ 
s\, y_L = k_L,  && s\, y_R =  k_R, \nonumber \\
s\, \overline{k}_L =-\overline{y}_L,  &&
 s\, {\overline{k}}_R =-{\overline{y}}_R, \nonumber \\ 
 s \,b_L = s\, k_L  = s\, \overline{y}_L =0, &&s \,  b_R =s \, k_R  
= s \,{\overline{y}}_R = 0,
\end{eqnarray}

These BRST transformations are nilpotent, 
$
 s^2 =0 
$, and they also commute with the quantum gauge transformations,
 $\delta_{qg}s = s\delta_{qg}$.
The total Lagrangian density   is  invariant 
under these  BRST transformations, $ s\, \mathcal{L}_t =0$, so 
 we can obtain  
a conserved  BRST charge  $Q_B$. This conserved charge  commutes with the Hamiltonian of the theory and 
generates the BRST transformations. We can also use this change to  
 project out the physical state and show that the $S$-matrix is unitarity. 
The states that are
 annihilated by $Q_B$ are defined to be the  physical states,
\begin{equation}
 Q_B |\phi_p \rangle =0. 
\end{equation}
The inner product of physical states, which are obtained 
by the action of $Q_B$ on unphysical states $|\phi_{up} \rangle $, vanishes with all
physical states,
\begin{equation}
 \langle \phi_p  |Q_B |\phi_{up} \rangle =0.
\end{equation}
This is because $Q_B$ is the generator of the BRST transformations and these transformations are 
 nilpotent,  and so we have
\begin{eqnarray}
 Q^2_B |\phi\rangle &=& 0.
\end{eqnarray}
Thus,  the relevant physical states 
actually are those that are not obtained by the action
of $Q_B$ on any other state.
Now we can write a $S$-matrix element as 
\begin{equation}
\langle\phi_{pa,out}|\phi_{pb,in}\rangle = 
\langle\phi_{pa}|\mathcal{S}^{\dagger}\mathcal{S}|\phi_{pb}\rangle,
\end{equation}
where the asymptotic physical states are given by 
\begin{eqnarray}
 |\phi_{pa,out}\rangle &=& |\phi_{pa}, t \to \infty\rangle,
 \nonumber \\
 |\phi_{pb,in}\rangle &=& |\phi_{pb}, t \to- \infty\rangle,
\end{eqnarray}
As $Q_B$   commute with the Hamiltonian, 
 the time evolution of any physical state will 
also be annihilated by  $Q_B$, 
\begin{equation}
 Q_B \mathcal{S} |\phi_{pb}\rangle =0.
\end{equation}  
This implies that the states $\mathcal{S}|\phi_{pb}\rangle$
 must be a linear combination of physical states $|\phi_{p,i}\rangle$. Thus, we can write 
\begin{equation}
\langle\phi_{pa}|\mathcal{S}^{\dagger}\mathcal{S}|\phi_{pb}\rangle
 = \sum_{i}\langle\phi_{pa}|\mathcal{S}^{\dagger}|\phi_{p,i}\rangle
\langle\phi_{p,i}| \mathcal{S}|\phi_{pb}\rangle.
\end{equation}
Now as the full $\mathcal{S}$-matrix is unitary
this relation
 implies that the  $S$-matrix restricted to
physical sub-space is also unitarity. 

\section{Conclusion}
In this paper we  analysed  the ABJM theory in $\mathcal{N} =3$ harmonic 
superspace formalism. The harmonic superspace variables used were 
 parameterized by the coset $SU(2)/U(1)$ and thus
had manifest $\mathcal{N} =3$ supersymmetry. 
The ABJM theory expressed in these variable had manifest $\mathcal{N} =6$
supersymmetry. 
We  analysed  
the quantum gauge transformations of this theory  
in 
gaugeon formalism. The gauge fixed ABJM  theory in the gaugeon formalism had extra fields called the gaugeon fields.
 This made the Hilbert space of the theory large enough to consider quantum gauge transformations. 
We  also analysed the 
BRST transformations of 
 this theory and used them to show that
the unitarity of the $\mathcal{S}$-matrix. 
These result could also have been  obtained 
by using
a conventional BRST  symmetry along with the Yokoyama's subsidiary 
condition. 
It is well know that for gauge theories there is symmetry  dual to the BRST  symmetry. 
This symmetry  is called the anti-BRST symmetry. 
It will be interesting to include the anti-BRST transformations in the present analysis. It is
expected that the results obtained will again be the same as 
obtained 
by using a conventional anti-BRST  
symmetry along with the Yokoyama's subsidiary 
condition.

Chern-Simons theories also  have important condense matter  applications. 
This is because the fractional quantum hall effect is generated by    
 Chern-Simons theories \cite{a}-\cite{d}. 
Fractional quantum hall effect is a property of a collective state in which electrons
 bind magnetic flux lines to make new quasi-particles. 
These quasi-particles have a fractional elementary charge.
Thus, the
fractional quantum Hall effect is based 
on the concept of statistical transmutation. 
In two dimensions, fermions can be described as charged bosons carrying
an odd integer number of  flux quanta. The fractional quantum hall effect is generated by    
 Chern-Simons theories
 fields coupled to these bosons.  If these electrons are analysed in a 
 combined external and statistical magnetic
field, then at special values of the filling fraction the statistical 
field cancels the external field,
in the mean field sense. At these values of 
the filling fraction and the system is described as a gas of bosons feeling no net
magnetic field. Thus, these bosons condense into a homogeneous ground state. 
This model also describes the 
 existence of vortex and anti-vortex excitations.

Lately, supersymmetric 
generalisation of the fractional quantum Hall effect has also been investigated \cite{a11}-\cite{a4}.
The physical properties of the topological
excitations in the supersymmetric quantum Hall liquid have also been discussed in a dual supersymmetric
Chern-Simons theory \cite{a5}.  Furthermore, the fractional quantum Hall 
effect is closely related to  noncommutativity  of the spacetime \cite{fqhfqh1}-\cite{fqhfqh2}. 
Thus, the results of this paper  can have interesting condensed matter applications. 
This is because we can analyse the supersymmetric fractional quantum Hall effect in the gaugeon formalism see 
if it generates some new physics. In fact, we can also include different deformations 
of the Harmonic superspace and then analyses this deformed ABJM theory in the gaugeon formalism. 
These deformations are expected to  
  change the behavior of  fractional condensates and thus
have  important consequences for the transport properties in the quantum hall  system.
  Recently,   supersymmetric
Chern-Simons theory  has defined to be
been used to study fractional quantum Hall effect via holography \cite{e}. 
In fact, the ABJM theory has   been used to study various interesting
 examples of $AdS_4/CFT_3$ correspondence \cite{6}-\cite{10}.  
 It will be interesting  to
 analyse similar effects in the harmonic superspace in gaugeon formalism. 

It may be noted that $\mathcal{N}=1$ supersymmetric 
Chern-Simons theory  supersymmetry coupled to
 parity-preserving matter fields 
has also been  
analysed using the Parkes-Siegel formulation \cite{1and1}.
In the Parkes-Siegel formulation the dimensional reduction
 from $(2+2)$ to $(2+1)$ of massive Abelian $\mathcal{N}=1$
super-$QED$ coupled to a self-dual super-multiplet produces 
 simple supersymmetric
extension of the $\tau_3QED$ coupled to  Chern-Simons theory.
It will be interesting to derive similar result in the harmonic superspace.

\end{document}